**Title:** Strain-engineered interaction of quantum polar and superconducting phases

**Authors:** Chloe Herrera[1], Jonah Cerbin[1], Kirsty Dunnett[2], Alexander V. Balatsky[1,2,3], and Ilya Sochnikov[1,*]

**Affiliations:**

[1] *Physics Department, University of Connecticut, 2152 Hillside Rd, Storrs, CT 06269*

[2] *Nordita, KTH Royal Institute of Technology and Stockholm University, Roslagstullsbacken 23, SE-106 91 Stockholm, Sweden*

[3] *Institute for Materials Science, Los Alamos, NM 87501, USA*

**One Sentence Summary:** Tension boosts superconductivity in strontium titanate via enhanced interaction between the incipient quantum polar and superconducting phases, explaining preformed pairing.

**Abstract:**

Much of the focus of modern condensed matter physics concerns control of quantum phases with examples that include flat band superconductivity in graphene bilayers (*1*), the interplay of magnetism and ferroelectricity (*2*), and induction of topological transitions by strain (*3*). Here we report the first observation of a reproducible and strong enhancement of the superconducting critical temperature, $T_c$, in strontium titanate (SrTiO$_3$) obtained through careful strain engineering of interacting superconducting phase and the polar quantum phase (quantum paraelectric). Our results show a nearly 50% increase in $T_c$ with indications that the increase could become several hundred percent. We have thus discovered a means to control the interaction of two quantum phases through application of strain, which may be important for quantum information science. Further, our work elucidates the enigmatic pseudogap-like and preformed electron pairs phenomena in low dimensional strontium titanate (*4*, *5*) as potentially resulting from the local strain of jammed tetragonal domains.

**Main text:**

Among the main goals of this work is to address the open question of the nature of the superconducting pairing mechanism in strontium titanate (STO) (*6*, *7*) and to inspire searches for enhanced superconducting temperatures in materials not just with suppressed to zero Kelvin structural transitions, as in (Ca$_x$Sr$_{1-x}$)$_3$Rh$_4$Sn$_{13}$ (*8*), MoTe$_2$ (*9*) and Lu(Pt$_{1-x}$Pd$_x$)$_2$In (*10*), but with *incipient* quantum phase transitions, for example, ScF$_3$ which has a structural quantum phase transition (*11*), and may become superconducting when doped (*12*). It has been predicted that superconducting doped strontium titanate with its peculiar phonon dynamics (*13*–*17*) is an example of a superconductivity arising near an incipient quantum polar (quantum ferroelectric) phase transition (*4*, *7*, *18*–*27*), but this has not been fully demonstrated experimentally, in part, due to the fact that existing results on isotope effect and Ca substitution (*25*, *28*) may be explained by non-uniformity in the chemical composition, and the absolute enhancement of the critical temperature values have not been found.

It is also unusual to find a pseudogap-like behavior in superconductors that cannot be explained by compositional inhomogeneities, as is the case in cuprates (*29*). A pseudogap-like behaviors, such as a tunneling gap and a 2*e* charge transport, occur in STO at temperatures up to about 0.9 K, almost twice the bulk superconducting transition temperature (*4*, *5*). This preformed pairing cannot be explained by local chemical variations as in inhomogeneous or granular systems (*30*) because doped STO does not superconduct at these temperatures (*31*).

We applied *in situ* uniaxial compressive and tensile strains – a clean non chemical control (*32–35*) – to narrow thin bars of SrTi$_{1-x}$Nb$_x$O$_3$ crystals (Nb-doped STO) with several doping levels along a specific crystallographic axis (**Figure 1 A-D**). The shape is chosen to promote the *x*-monodomain state (*36*) (large *x*-domains along the force direction, and long side of the samples) in the low temperature tetragonal phase. In this controlled geometry, we can tune the crystals towards the polar quantum phase transition with uniaxial tensile strain which softens the ferroelectric mode (*7*, *37*) while imaging the remaining tetragonal domains and measuring the resistive transition from which we determine $T_c$. To our knowledge, no prior uniaxial *tensile* strain experiments in *monodomain* samples exist for superconducting STO (*32*).

Our experimental setup consists of a custom-built strain cell, and a polarizing optical microscope installed in a dilution refrigerator, see **Figure 1 C** and the Supplementary Materials (SM). We used very low excitation currents between 100 nA – 100 µA and an ultra-low noise amplifier and resistance bridge from Lakeshore to measure the resistive phase transition, with hundreds of pico-volts signals (further technical details are in the SM).

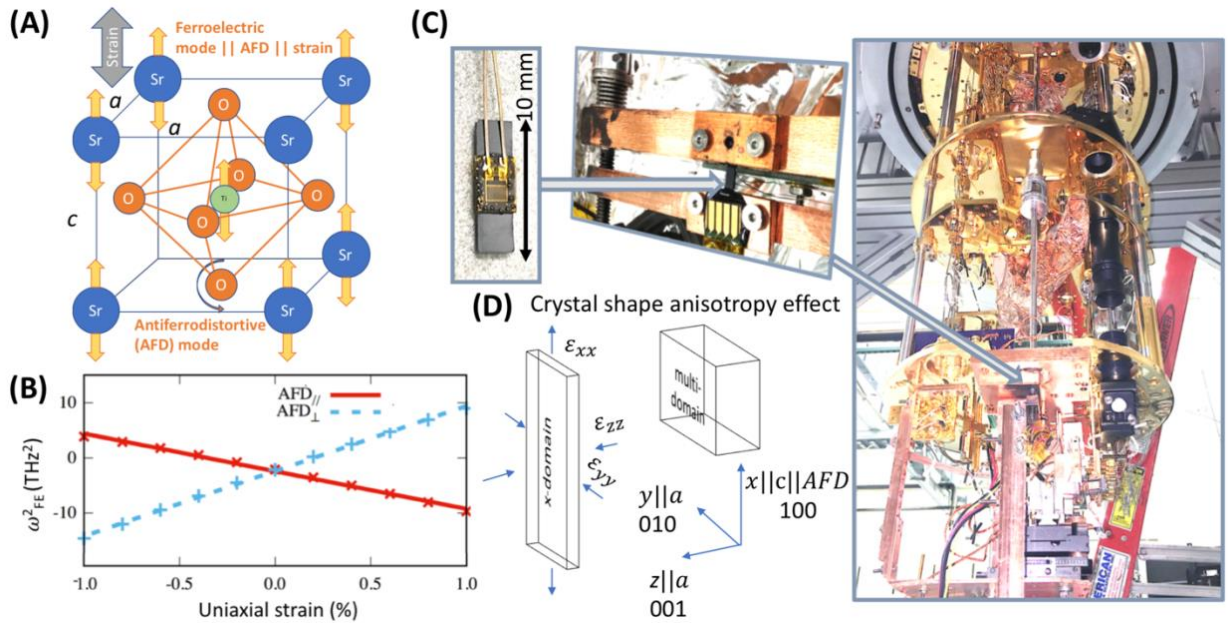

*Figure 1. STO ferroelectric soft modes and experimental setup for probing resistance and optical imaging under strain. (A) Approximate atomic displacements of the ferroelectric phonon mode. (B) Calculated softening of the ferroelectric phonon mode parallel to the antiferrodistortive (AFD) rotation axis under uniaxial tensile strain (based on Ref. (7)). (C) Long, narrow and thin samples (typically 10x2x0.27 mm$^3$) were clamped in a custom-built strain cell to apply tunable strain along the long side of the crystals. Electronic transport and optical measurements were performed in situ while the strain was continuously tuned in a closed cycle dilution refrigerator (right). (D) Strain in the x direction, $\varepsilon_{xx}$, measured by the gauge has the opposite sign to the z and y strains, $\varepsilon_{zz}$*

and $\varepsilon_{yy}$, which can be determined using the Poisson's ratio (7). Further experimental details are in the SM.

Our main finding is that the critical temperature of SrTi$_{1-x}$Nb$_x$O$_3$ increases strongly and reversibly under tensile strain (**Figure 2**) in crystals with preferentially oriented domains (**Figure 3**). The increase was observed in many samples (see **Figure 4 A-D** for typical data). Lower doping samples **Figure 4 A** showed higher crystal fragility, limiting the strain that we could induce (*38*). We could reach higher strains in samples with higher Nb content. Overdoped samples (e. g. **Figure 4 D**) did not show a substantial increase in $T_c$. Therefore, the highest $T_c$ is reported for the nearly optimally doped samples (**Figure 4 B** and **C**).

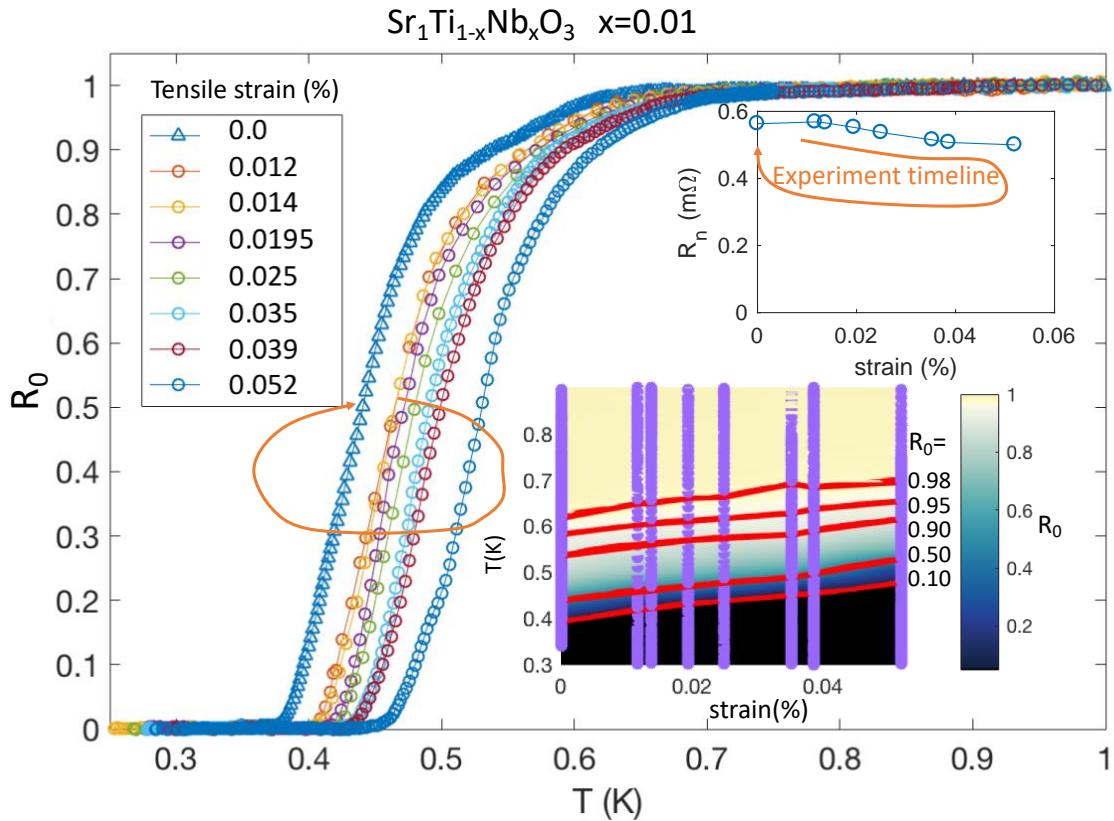

*Figure 2. Large reversible increase in the critical temperature of Sr$_1$Ti$_{0.99}$Nb$_{0.01}$O$_3$ under uniaxial tensile strain that drives the sample towards the quantum phase transition. The curves are normalized $R_n=R(1K)$ where the resistance is weakly temperature dependent: $R_0=R(T)/R_n$. The inset shows $R_n$ with the value returning to the original value after the release of strain (arrows follow the timeline of the experiment). The reversibility demonstrates that the critical temperature change is not driven by the formation of lattice defects. The substantial shift in the resistive transition towards higher temperatures (see inset for the transition plotted as red colored isolines of $R_n$) means that $T_c$ changes cannot be explained by small Fermi surface modifications (39), but are related to the phonon spectrum modification.*

We also find that the critical temperature of SrTi$_{1-x}$Nb$_x$O$_3$ decreases under compressive strain, consistent with previous studies on multi-domain samples (*32*). However, one needs to be cautious as in multi-domain samples $T_c$ changes could depend strongly on the domain orientation (*32*) and therefore cannot be directly compared to our results. The aspect ratio change of our samples with cooling from 300 K to below 4 K is in strong agreement with the scenario where the *c*-axes of most of the domains in the sample are along the strain direction (**Figure 3**). We also show supporting optical microscopy images in the SM. The strain gauge and the optical measurements imply that *y*-domains and *z*-domains (*40–43*) did not dominate the supercurrent path and were not critical for observed changes of the transition temperature. The preferential orientation of the tetragonal domains relative to the strain direction was a necessary condition for observing the enhancement of $T_c$.

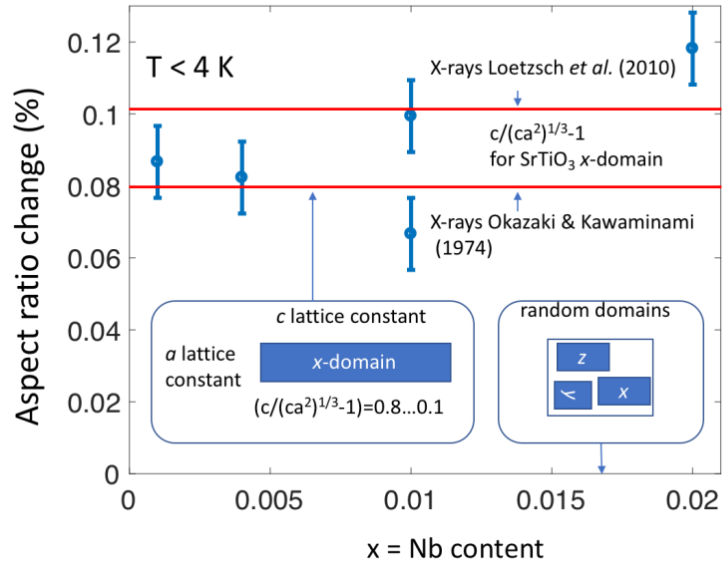

*Figure 3. Strain gauge measurements show preferential direction to the tetragonal domains in the measured crystals along the applied strain. Ordinate: the aspect ratio change in long samples relative to the thick square samples (**Figure 1 D**) upon cooling below the 110K cubic-tetragonal transition (44–46). The data (blue circles) are for the five samples in **Figure 2** and **4** (x=0.01 sample from **Figure 4 C** shows higher aspect ratio than sample in **Figure 2**). More details on strain gauges and complimentary optical images are in the SM.*

The calculated frequencies of the phonon modes parallel and perpendicular to the antiferrodistortive (AFD) c-axis are shown in **Figure 1 B**. Under tensile strain the mode with atomic vibrations parallel to the AFD axis softens – frequency decreases (DFT does not fully account for the quantum paraelectric fluctuations but shows the trends in the mode softening correctly). Comparing our data with the calculated ferroelectric phonon modes indicates that the increase in $T_c$ is correlated with the softening of the ferroelectric phonon mode branch parallel both to the AFD axis and the applied strain (**Figure 1**). The decrease in $T_c$ is correlated with the hardening of the same ferroelectric phonon mode.

Meanwhile the ferroelectric phonon modes perpendicular to the AFD (and strain) axis harden under tension and soften under compression (*47*), and it is a reasonable assertion that these modes participate in pairing. Under isotropic pressure, all the ferroelectric phonon modes harden (*7*), and $T_c$ decreases (*27*, *32*). Therefore, it appears that the phonon mode parallel to the AFD axis is the one that is most strongly coupled to the superconducting order, in agreement with one of the proposed theoretical scenarios (*7*). The reported here *asymmetric* behavior under bidirectional strain indicates that the critical temperature is not determined by just any softest ferroelectric mode but linked strongly to the specific mode parallel to the AFD axis, providing an important insight into the superconducting coupling mechanism of strontium titanate.

The divergence-like feature in $T_c$ at high strains in **Figure 4 B** and **C** deserves special attention. In the model of ferroelectric phonon mediated coupling considered here (*7*, *24*), $T_c$ is expected to grow sharply as the phonon mode frequency decreases on the approach to the incipient quantum phase transition. It is very likely that the sharp upturn in $T_c$ in our data is exactly the start of the singularity that would occur if the quantum phase transition occurred near these carrier concentrations. Another interesting observation is that the overdoped samples (**Figure 4 D**) showed a very weak response to strain even for relatively large strains, likely because the overdoped samples are farther away in the phase space from the expected polar quantum critical point (see the model in Ref. (*7*) and SM).

Although we found robust evidences that both compressive and tensile strains probed predominantly the domains oriented along the applied strain direction some contribution from multi-domains is not fully ruled out. $T_c$ increased and diverged upon tensile strain, but the exact rates of change at lower strains varied between measurements. This variance may be related to the exact current distributions, small multi-domain patches, or non-uniform strain distributions. In particular, these effects may be present near the leads and mounting edges of the samples. Despite these possible factors, we emphasize that the suppression of $T_c$ under compressive strain and the growth and sudden increase of $T_c$ (divergence) under tensile strain were reproduced in multiple samples as summarized in **Figure 4 E**.

**(A)**

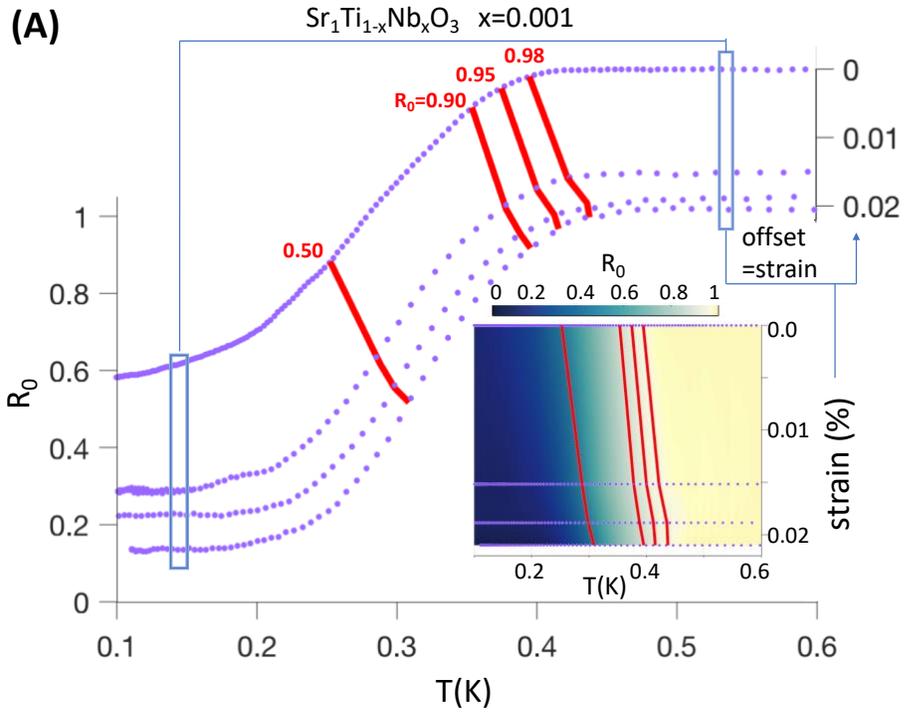

**(B)**

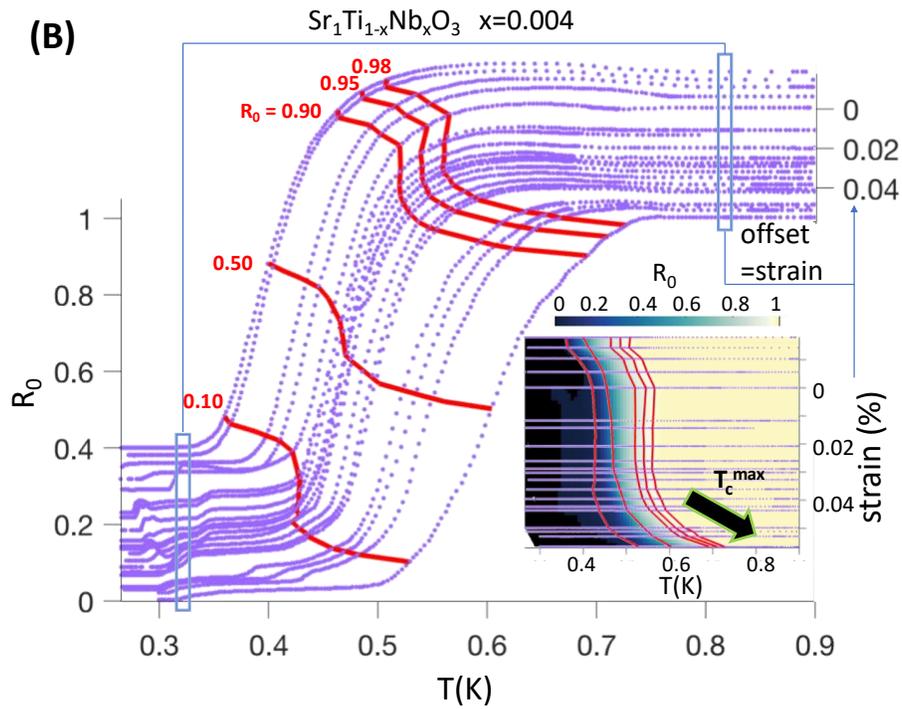

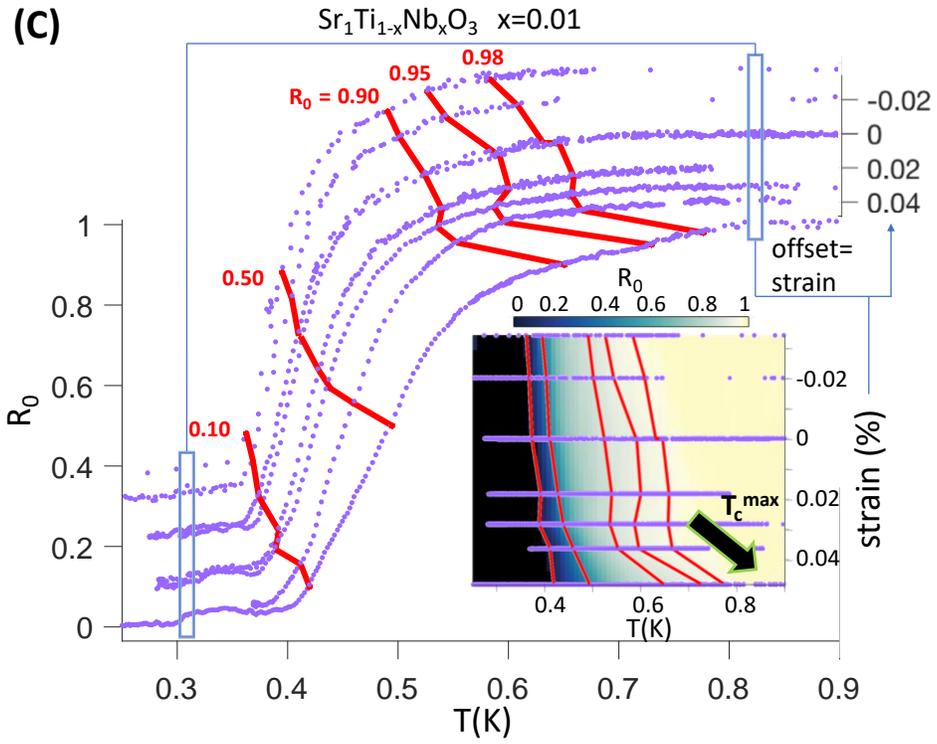

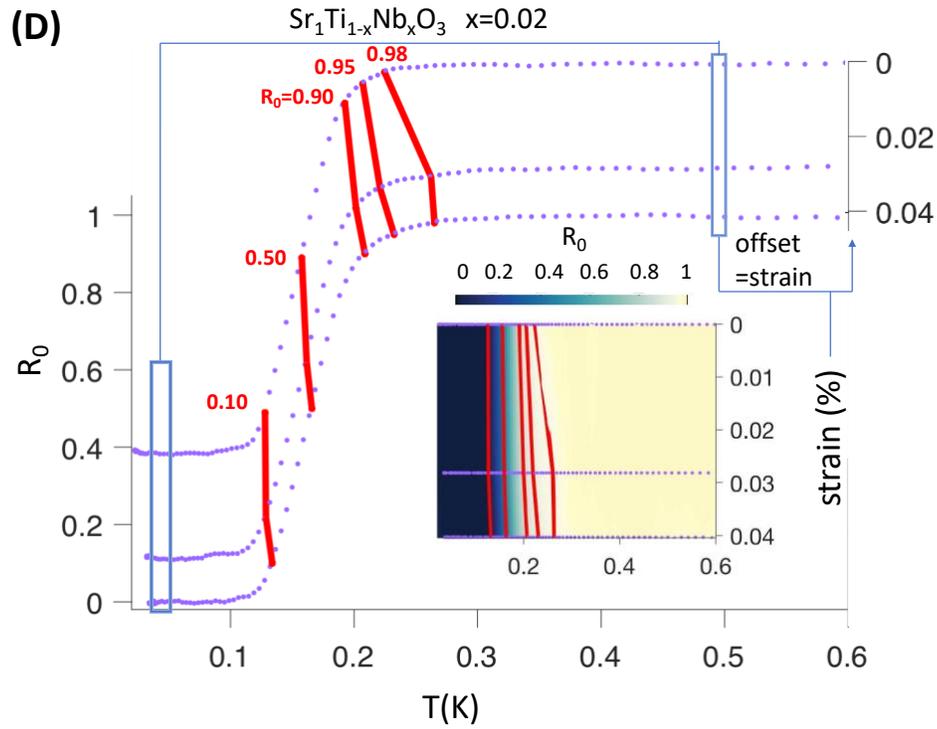

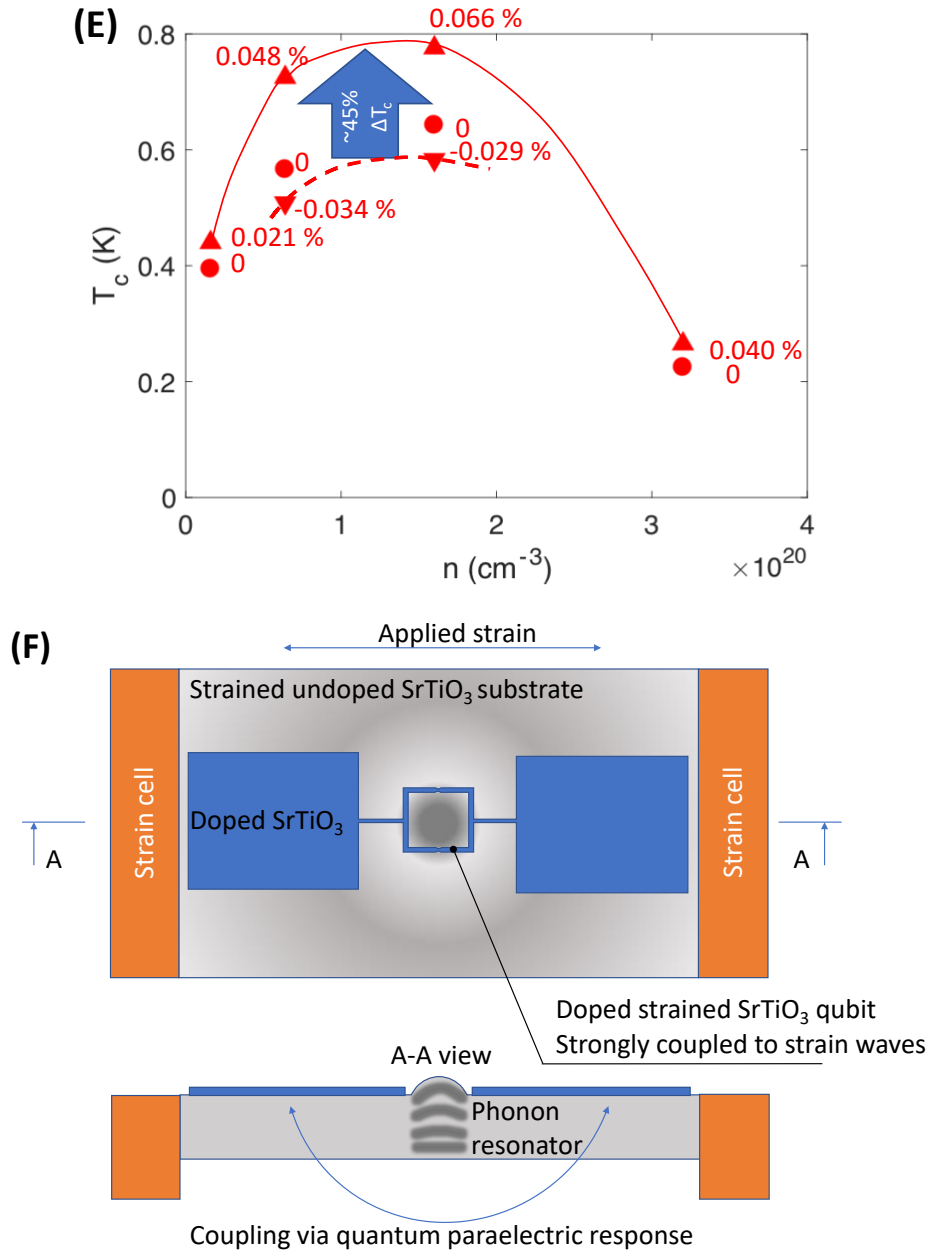

*Figure 4. Huge increase in the onset of the critical temperature and general enhancement of $T_c$ across the phase diagram. (A)-(D) Representative data for $Sr_1Ti_{1-x}Nb_xO_3$ crystals with four different doping levels. The critical temperature decreases under compressive and increases under tensile strains. Near optimally doped samples (B) and (C) showed a diverging rise in $T_c$, while (many) underdoped samples, e. g. the one shown in (A), split (usually normal to the strain axis) at ~0.021% strain. No limit is observed for the maximal possible critical temperature, $T_c^{max}$, as it appears that an onset of divergence in $T_c$ is seen as indicated by black arrows in (B) and (C). Resistance curves are offset for clarity. Note that in sample in **Figure 2** we have not reached high enough strains to see the divergence-like behavior (within the error bars of ±0.01 %). Overdoped sample in (D) and other overdoped samples did not show a substantial increase in $T_c$, consistent*

*with the discussed theoretical model (7). **(E)** $T_c$ as a function of doping. Upwards (downwards) pointing triangles – highest (lowest) $T_c$ at labeled strains. $T_c$ is defined as 98% of the normal resistance at 1K, i. e. the onset of the superconducting transition, which is more robust against the excitation current influence as explained in the SM. An increase of about 45% in $T_c$ occurs under the highest strains at near optimal doping level. The width of the transition either increased moderately or remained largely unaffected by strain. Error bars in determining the absolute strain are $\pm 0.01\%$. The red dashed and the black guide-to-the-eye lines fence the region of enhanced $T_c$ (note, these are not isostrain lines). **(F)** A proposed strontium titanate quantum sensor (transducer) for detecting coherent acoustic phonons with a superconducting qubit. The qubit-phonon coupling is based on the high sensitivity of superconducting strontium titanate to local lattice deformations as it is tuned towards the polar phase transition. An acoustic phonon could be delivered to the strontium titanate qubit which then changes its impedance as the Josephson energy of the strained system is higher than the unstrained system.*

The multi-domain aspect of STO samples is relevant to our suggestion that if the tetragonal domains are pinned or the sample is not completely freely mounted, relative local elongations on the order of 0.08-0.09 % (**Figure 3**) can occur in a staggered matrix of domains. It is plausible then to expect that some regions of the sample will effectively experience tensile strains of the same order as we have applied externally (approximately $\pm 0.05\%$). This can easily lead to an enhanced $T_c$ in microscopic regions or even along a percolative path, which can also happen at interfaces (*4*, *5*): the pseudogap-like phenomena (or pseudogap-like features) may therefore be related to strain inhomogeneity.

The asymmetric response of $T_c$ to strain also rules out the localized defects-associated phonons scenario of superconductivity proposed by Gor'kov (*48*) as the response to strain from random defects is expected to be symmetric. This significantly reduces the range of relevant theories of superconductivity in STO.

Further experimental efforts motivated by our results can focus on direct phonon measurement and manipulation techniques; combined with probes like electronic transport, scanning SQUID imaging (*49*), and nanoscale strain control (*42*), phonon manipulation could bring to life new ideas such as probing very high *dynamic* strain responses, potentially leading to even higher critical temperatures not accessible with the static strains used in this work.

The almost 0.8 K onset critical temperature reported here makes strained superconducting strontium titanate comparable to superconducting aluminum. Similar to aluminum, strontium titanate could be used in superconducting qubits. Hybrid structures of doped and undoped strontium titanate may bring new functionality to qubits. In **Figure 4 F**, we propose a superconducting-phononic qubit (*50*, *51*) that can transduce phonon vibrations in the quantum paraelectric medium to electromagnetic signals. We suggest that such devices will be more sensitive to phonons than ones based on conventional superconductors (*50*), since the impedance of the qubit will be very sensitive to elastic deformation near the divergence in $T_c$, the quantum paraelectric state of the undoped STO substrate will screen undesired charge fluctuations and the strong coupling may also allow smaller device sizes and better scalability.

We have reported on the discovery of continuously *tunable* interactions of entangled quantum polar and superconducting phases in STO which can explain a range of phenomena: from the dome of the superconducting critical temperature, to ruling out some roles of defects, and the pseudogap behavior.

We observe a sharply rising (divergent-like) superconducting critical temperature projecting no apparent cap on the critical temperature of STO as it is tuned towards the polar quantum phase transition.

**Acknowledgements:** We thank J. Hancock, J. Levy, Y. Kedem, N. Spaldin, J. Budnick and B. Wells for helpful discussions, J. Sheldon, B. Hines, and A. Jayakody for help with setting up the experiments, room temperature sample characterizations, and B. Willis for use of his equipment to make leads to the samples. **Funding:** The State of Connecticut, the Physics Department, and the Vice Provost for Research at the University of Connecticut provided financial support through research startup funds, and the Scholarship Facilitation award. The College for the Liberal Arts and Sciences at the University of Connecticut awarded C.H. a graduate student financial award. The work of J. C. was supported by the Mark Miller award. A.V.B. and K.D. were supported by US BES E3B7, VILLUM FONDEN via the Centre of Excellence for Dirac Materials (Grant No. 11744) and by Knut and Alice Wallenberg Foundation (2013.0096).

**Authors Contribution:** C.H., J. C. and I. S. planned and performed experiments, completed data analysis and interpreted the results. K. D. and A. V. B. took part in planning and interpretation of the experiments. All authors contributed to the manuscript preparation. **Competing Interests:** The authors declare no competing interests. **Data and materials availability:** All data are available as figures in the text and in the SM. **Materials & Correspondence**: ILYA.SOCHNIKOV@UCONN.EDU**Supplementary Materials:**

Supplementary Text

Figures S1-S5

Supplementary Materials References

# Supplementary Materials for

# Strain-engineered interaction of quantum polar and superconducting phases

Chloe Herrera, Jonah Cerbin, Kirsty Dunnett, Alexander V. Balatsky, and Ilya Sochnikov

Correspondence to: ilya.sochnikov@uconn.edu

**This PDF file includes:**

Supplementary Text

Figures S1 to S5

**Other Supplementary Materials for this manuscript include the following:**

Supplementary Materials References

**Supplementary Text**

1. Samples

Experiments on single crystals of SrTi$_{1-x}$Nb$_x$O$_3$, with nominal *x* values of 0.01, 0.004, and 0.001, are reported here. They were purchased from Furuchi Inc, Japan, and are similar in the quality to crystals used in other studies (*20, 21*). The (1 0 0) crystals were typically polished to an optical quality and then cut into 2x10x0.27 mm$^3$ rectangular pieces with the long side in the (1 0 0) crystallographic direction, which becomes the tetragonal c-axis (AFD axis) upon cooling through the 110 K structural phase transition. The high quality of the crystals is confirmed by the very high R(300K)/R(1K) ratio (around three orders of magnitude, depending on the doping level).

2. Strain measurements

We attached strain gauges SGT-1/350-TY11 from Omega Inc. to the reverse side of the samples using Stycast epoxy before mounting in the strain cell. The resistive strain gauge arrangement allowed us to verify the *residual* strains in the as-cooled samples (e. g. due to imperfections in compensations for thermal shrinkage of the clamp or gluing, which occurred on a scale of ± 1 μm). This has been done at temperatures below 1K where we find the as cooled strain to be limited to about ±0.01 %, which gives the accuracy of the absolute value of the measured strains, while the relative accuracy is much higher for a given sample or cool-down. We were not able to cool samples with a controlled non-zero applied strain through the 110K transition as the sample formed permanent dislocations if cooled with applied force possibly due to softening of a structural mode at the R- point of the Brillouin zone (*52*).

3. Strain cell

**Figure S 1** shows our thermally compensated strain cell, that could accommodate the changes in sample size and aspect ratio, and with a strain measurement accuracy of about ±0.01 %. We verified the accuracy by comparing the gauge readings on clamped samples to the readings of the strain gauges mounted on free standing samples.

We mounted the samples with attached gauges in the copper strain cell using Stycast epoxy, leaving about 7 mm of the sample length for the probing of superconducting properties, and optical imaging. The strain cell had a lead screw that was controlled by a high torque, low speed geared DC motor. The lead-screw driving rod was fed through the entire height of the Bluefors LD-250 dilution refrigerator (**Figure S 2**), with thermalization points at every cold plate, bronze bearings at some low temperature stages, and a high-vacuum ferrofluid rotational feedthrough at the room temperature stage.

When cooled from room temperature to as low as 20-50 mK, the samples experience at most ±0.01 % of residual strain. As-cooled samples therefore typically had very close to zero strain (compared to free standing samples). This design provides small strains and minimizes dislocation defects in the sample.

This performance metric of our strain device is comparable to piezo-based devices (*31, 32*), but our copper clamp is essential for efficient thermalization at dilution refrigerator temperatures. We chose to develop an all metal strain rig (lead screw clamping), because of better control for tensile strain, and the ability to apply very large forces (~1000N for tensile and larger for compressive strains). These forces translate into more than 2GPa of stress on samples which allows us to reach any strain within the elastic deformations range of STO. In our clamp, we could attach resistive gauges directly to large samples for *in situ* strain monitoring.

4. Optical monitoring of tetragonal domains in a dilution refrigerator

We constructed a polarizing optical microscope in our BF-250 dilution refrigerator using mostly conventional optical components with some care of mounting the lenses and the mirrors in as strain-free a manner as possible. Some lenses simply rested on stopping rings inside aluminum lens tubes to avoid cracking due to thermal contraction. We degreased and removed plastic parts from Thorlabs and Newport opto-mechanical components to make them more vacuum and cryogenics compatible.

The microscope has a conventional geometry (**Figure S 2**). An LED source is located outside the cryostat. An 800 micrometers diameter silica HV vacuum-compatible optical fiber carries the light. A polymer film with no glass layers polarizes the light and additional lenses refocus it while passing to a 50:50 beam splitter in a Köhler illumination scheme. We installed a series of infrared-opaque windows at each cold plate of the cryostat. The light focuses on the sample with an objective lens (with two adjustable rerouting mirrors), it reflects back from the sample, passes through the beam splitter, a condenser lens, another polarizer, and through a viewport to a 15 MB CMOS camera detector. The camera acquired the images, while additional optomechanical components aligned the camera outside the cryostat. A representative series of optical images at a range of strains is shown in **Figure S 3**. We obtain a spatial resolution of 5-8 micrometers from the microscope, mainly depending on the numerical aperture (NA) of the objective lens (various lenses were used during this study). The microscope could not resolve some domains which can be smaller than 0.5 micrometers: This microscope design is a compromise between the very wide field of view (6 mm diameter) and the spatial resolution.

When the samples undergo the antiferrodistortive phase transition around 110K, large *x*-domain regions form along the long side of the sample due to the shape anisotropy as confirmed in our *in situ* images (see **Figure S 3** showing tensile strain series). Relative stability of the preferential *x*-domain orientation under applied strains is a result of the shape anisotropy of our samples, consistent with previous reports (*33*). The effect is not observed in other geometries (*52*).

In addition, as presented in **Figure 3** of the main text, we monitored the gauge reading at temperatures in the range of 300K - 20mK and found it to be consistent with the elongation of the sample (or more accurately, change in the aspect ratio) in the tetragonal phase on the order of 0.09% consistent with the direct lattice constants measurements (*41-43*).

5. Electronic Transport Methods

We measured the resistance of samples using a Lakeshore 372 AC resistance bridge which also served as the temperature controller. The resistance bridge also included a Model 3708 8-channel preamplifier and scanner. When combined with the resistance bridge, this gave an exceptionally low voltage noise floor of just 2 nV$_{RMS}$/Hz$^{1/2}$. When combined with filtering, hundreds of pico-volt signals could be measured (**Figure S 4**).

The resistance vs. temperature curves (**Figure S 4**) show that there is a substantial contribution from vortex or phase slip dynamics (*53*) in the middle of the transition, a phenomenon not previously studied in detail in STO. Further, investigation of vortex dynamic effects using tools like scanning SQUID microscopes will be worthwhile pursuing (*54*).

One of the implications of the observed current dependence is that the accuracy of comparison of published data sets might be compromised or complicated by the differences in the excitation currents and the specific criteria used to define the critical temperature. We suggest that the 98% onset of the normal resistance threshold is a more robust way of determining $T_c$ than anywhere in the middle of the transition as it is less sensitive to the excitation current. Only limited full R(T) data is available in the literature, and sometimes excitation current densities are not known, this is why in **Figure 4 E** we compare our measurements to one work where we were able to determine the onset $T_c$. We intentionally avoided comparing $T_c$ as would have been defined using 10%, 50% or 90% of R$_n$ criteria, because the previous data and our results are not always acquired using the same excitation current densities.

## 6. $T_c$ on the approach to ferroelectricity

The pairing mechanism in strontium titanate has not, to date, been determined unambiguously, and several origins have been suggested (*24*, *55–57*). One proposal is that, due to the incipient ferroelectric nature of STO, and the proximity to the ferroelectric quantum critical point (FE QCP), the ferroelectric phonon modes, which soften (decrease in energy) on tuning towards the FE QCP, could provide the superconducting pairing.

The model of Refs. (*7*, *24*) is one particular example that leads to a large, divergent, increase of $T_c$ when a ferroelectric phonon mode, assumed responsible for pairing, softens on approach to the FE QCP, although other models could be proposed with similar features. The critical temperature is given by BCS theory with energy scale $\epsilon$:

$$T_c = \epsilon e^{-1/\lambda}$$

With Eliashberg strong coupling:

$$\lambda = \int_0^\infty d\omega_q \frac{\alpha^2(\omega_q)}{\omega_q} F(\omega_q).$$

The simplifications of $q = 0$ and constant $\alpha^2(\omega_q)F(\omega_q)$ capture the key feature of a van Hove singularity and large increase in $T_c$ near the FE QCP where $\omega_{q=1} \to 0$ (*24*). The squared phonon energies are linear with strain $u$, so $\omega_0^2(u) = \omega_0^2(0) + bu$ for strain tuning towards the FE QCP (*7*).

In **Figure S 5**, $T_c(u)$ and $\omega_0^2(u)$ are plotted for a single carrier concentration. At small strains there is a slow change in $T_c$ which increases rapidly as the FE QCP is reached. The green shaded area indicates the region probed by this experiment which does not drive the STO samples to becoming ferroelectric. In the ferroelectric phase an extended description is needed, such as that developed in Ref. (*26*).

### Normal state resistance's relation to changes in $T_c$

Doped $SrTiO_3$ is probably an s-wave superconductor, meaning that $T_c$ is fairly insensitive to small changes in disorder (*58*). We found no statistically significant changes in the normal state resistance, $R_n$ (**Figure S 6**) to indicate a connection between $R_n$ and $T_c$. Under strain, $R_n$ could either increase or decrease in different samples with the same Nb content (e. g. $R_n$ of samples in **Figure 2** and **Figure S 6 A**). The changes in $T_c$ that we have reported cannot be explained as originating in defects created as a result of the induced strain. Similarly, the changes in the effective masses of carriers and Fermi energies are not large enough (*36*) to explain the rise in $T_c$. We therefore conclude that the enhancement in $T_c$ originates in the polar phonon changes in the material in agreement with our main hypothesis.

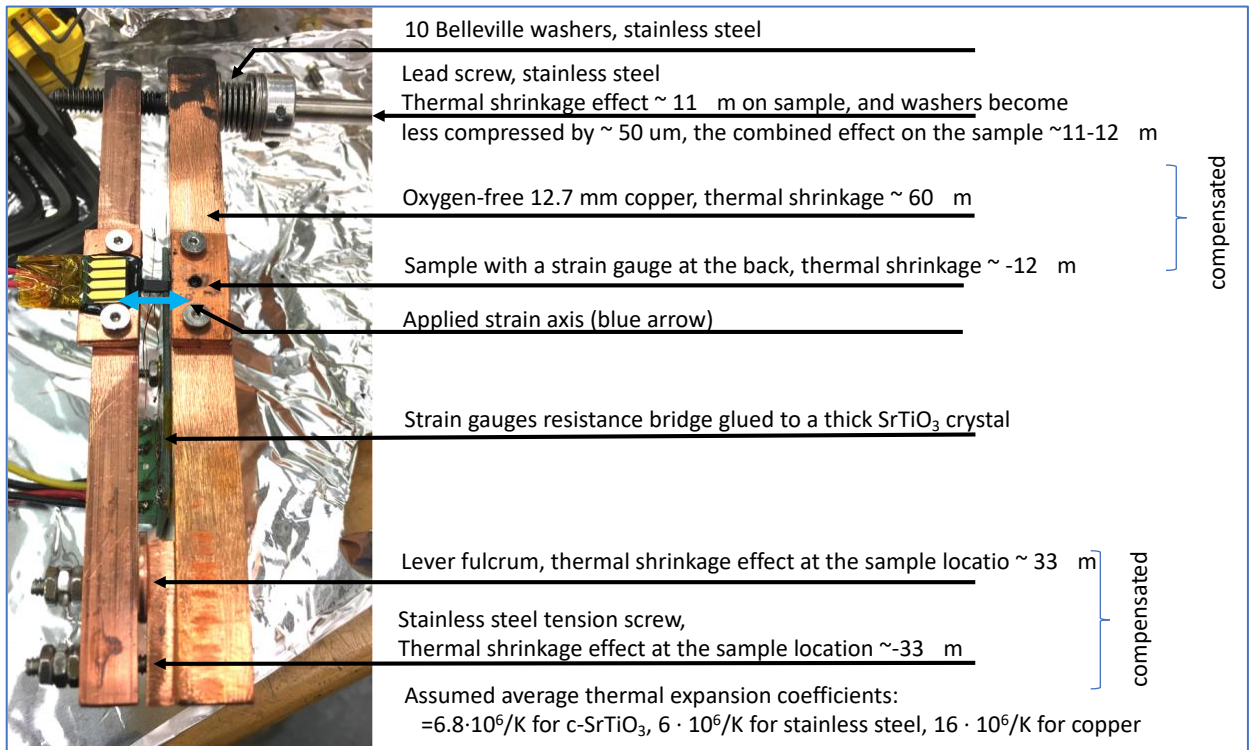

**Figure S 1. The bidirectional strain (compressive and tensile) cell.** *The geometry of the clamp, combined with the copper and stainless-steel parts compensates for the different thermal shrinkages of STO samples and the adjacent copper parts. The main material is oxygen free copper for use in a dilution refrigerator. The careful cell design and its simplicity are critical to the success of the measurements of samples which are essentially free of residual strain when cooled to temperatures of 10s of mK.*

Labels in figure:
- 10 Belleville washers, stainless steel
- Lead screw, stainless steel. Thermal shrinkage effect ~ 11 μm on sample, and washers become less compressed by ~ 50 um, the combined effect on the sample ~11-12 μm
- Oxygen-free 12.7 mm copper, thermal shrinkage ~ 60 μm
- Sample with a strain gauge at the back, thermal shrinkage ~ -12 μm
- Applied strain axis (blue arrow)
- Strain gauges resistance bridge glued to a thick SrTiO$_3$ crystal
- Lever fulcrum, thermal shrinkage effect at the sample location ~ 33 μm
- Stainless steel tension screw, Thermal shrinkage effect at the sample location ~ -33 μm
- Assumed average thermal expansion coefficients: $\alpha$ = 6.8·10$^6$/K for c-SrTiO$_3$, 6·10$^6$/K for stainless steel, 16·10$^6$/K for copper

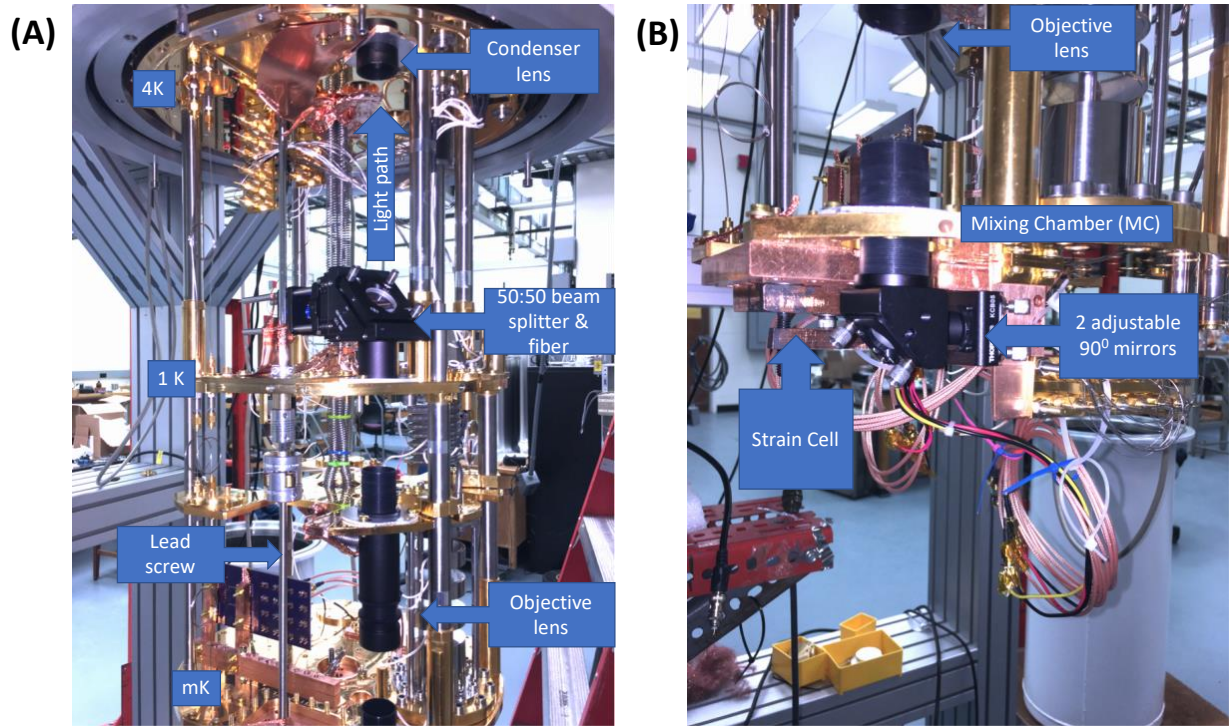

*Figure S 2. Polarized optical microscopy in the Bluefors dilution refrigerator to image the strained samples in situ.* **(A)** is the upper part of the experiment showing the optics of the detection system while **(B)** is the lower part and the view angle is from slightly to the right of (A). The horizontal plates are different temperature stages as indicated. The sample is located behind the 90° adjustable mirrors in (B).

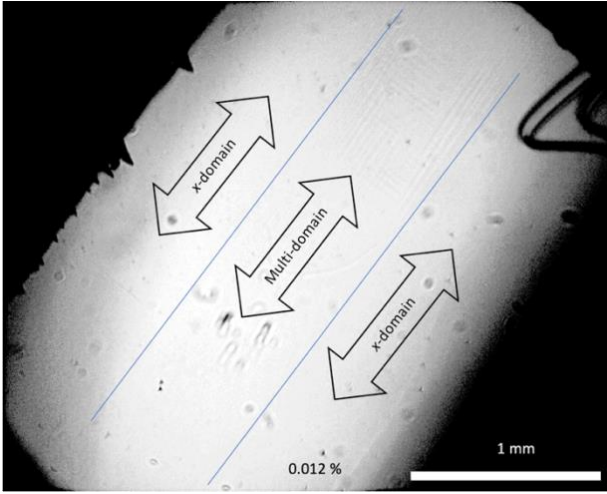 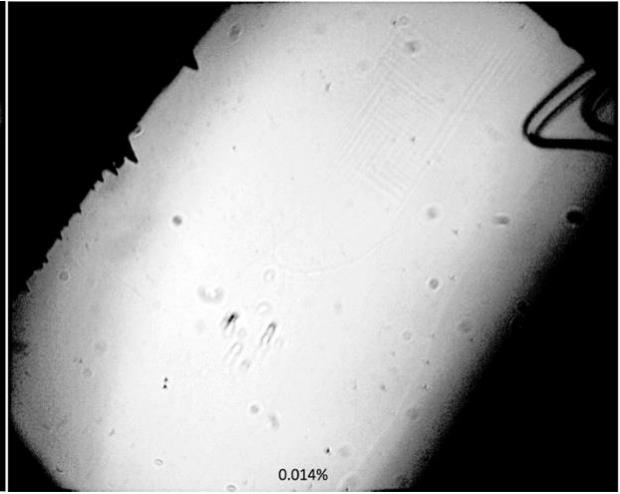
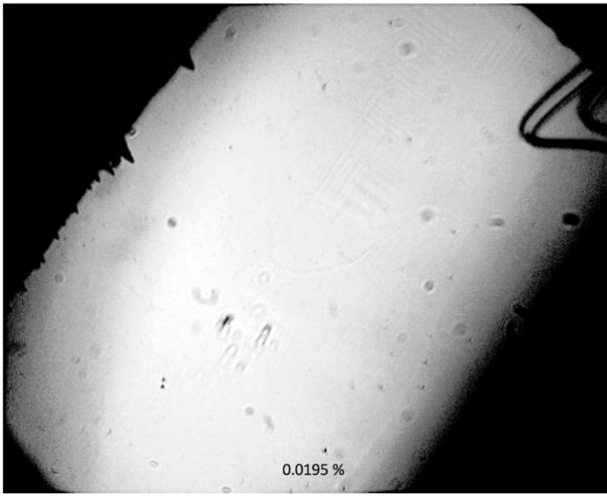 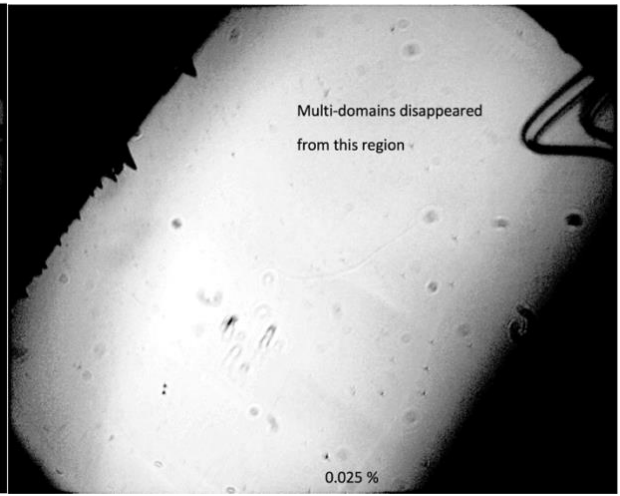
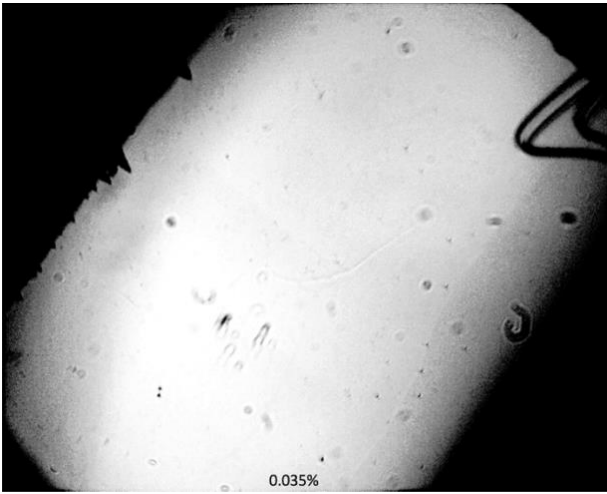 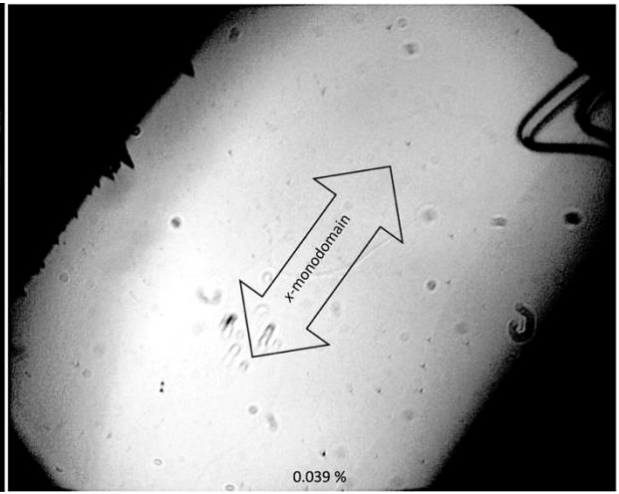

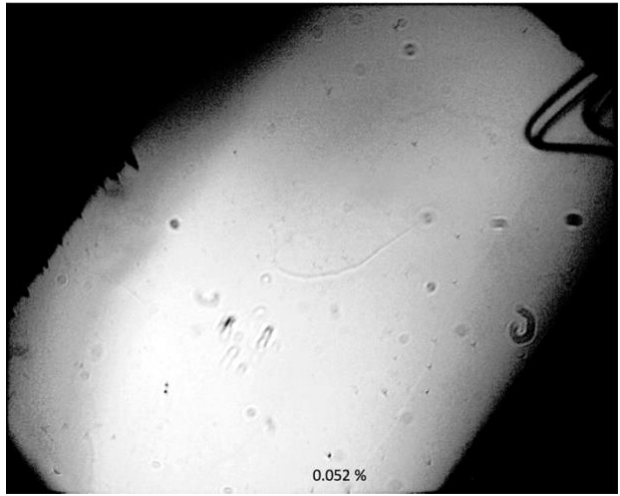

***Figure S 3. Optical microscope images of sample from Figure 2 (of main text) show primarily an x-domain state at low strain and gradual detwinning until a full x-domain state is achieved.** Images are taken simultaneously with strain tuning while the sample temperature stayed below 2K between resistance measurements. Optimally doped samples are not transparent, meaning that we detect surface tilts between 0º and 90º domain boundaries (the angles are relative to the (1 0 0) direction in the [1 1 0] plane). The 45º domain boundaries are not visible, as they are not accompanied by a surface change. The polarized optics was not effective in detecting the 45º boundaries. As strain is increased, we observed detwinning of the remaining multidomain part and the ends of the samples. The probing wires (one is visible as a bent feature in the top right of the images) were usually attached close to the long sides of the samples, where the x-domains are formed even for zero strain. Upon compression, we found that dislocations (35) occurred at strains lower than those required to switch the samples into states with the AFD axis perpendicular to the applied strain (data not shown). In the images, the tetragonal domains are only identifiable at the two or three lowest strains and disappear for larger strains.*

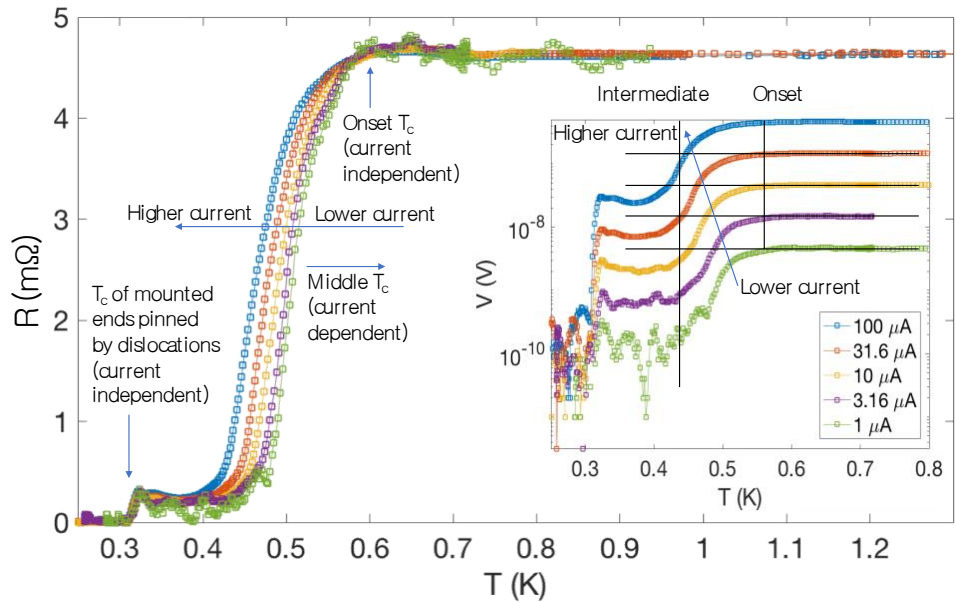

*Figure S 4. The transition temperature onset does not depend substantially on the excitation current while the middle of the transition does.* *The data shown is for x=0.004. 'Pinned' $T_c$ in sample regions near the mounting edges demonstrates that the dependence is not due to heating effects, but thermally activated vortex dynamics or phase slips. The noise in the 1 μA curve above the onset $T_c$ is ~200 pV (see inset for voltages at lower temperatures), corresponding to about three hours of acquisition for the entire curve and about 3 min of averaging per each data point. This noise level is in agreement with the internal noise-floor of the amplifier – any external noise is efficiently shielded. Following these current dependence results, we defined $T_c$ (**Figure 4 E**) as the onset of the drop in the resistance, which is, for practical purposes, determined as a temperature at which 98% of the normal resistance is reached.*

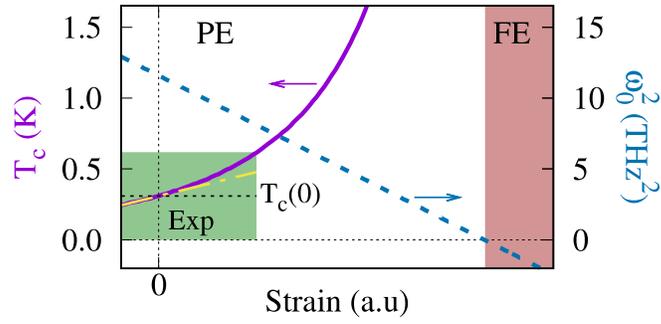

*Figure S 5. Theoretical diverging superconducting transition temperature:* *($T_c$ solid purple line) and squared phonon mode frequency (dark blue, dotted line) for carrier concentration $n_c$ = 6.25 × $10^{19} cm^{-3}$, $T_c(u=0) \approx 0.31K$. The green shaded region, up to $T_c = 2T_c(u=0)$, in which the yellow dash-dotted line is a linear fit to the $T_c$ under compression, highlighting the non-linear increase of $T_c$ under tension, covers the experimental region. In the model developed in Refs. (24) and (7), used here, the unstrained $T_c$ is fitted to the data of Ref. (19) The strain axis is a.u. because the calculation is done for isotropic strain which is also representative of phonon modes parallel to the AFD axis under uniaxial strain; the exact strain values for the experimental set up are not known in the model. Ferroelectricity (red shaded region, $\omega_0^2 < 0$) occurs for much larger strains than those required to double $T_c$.*

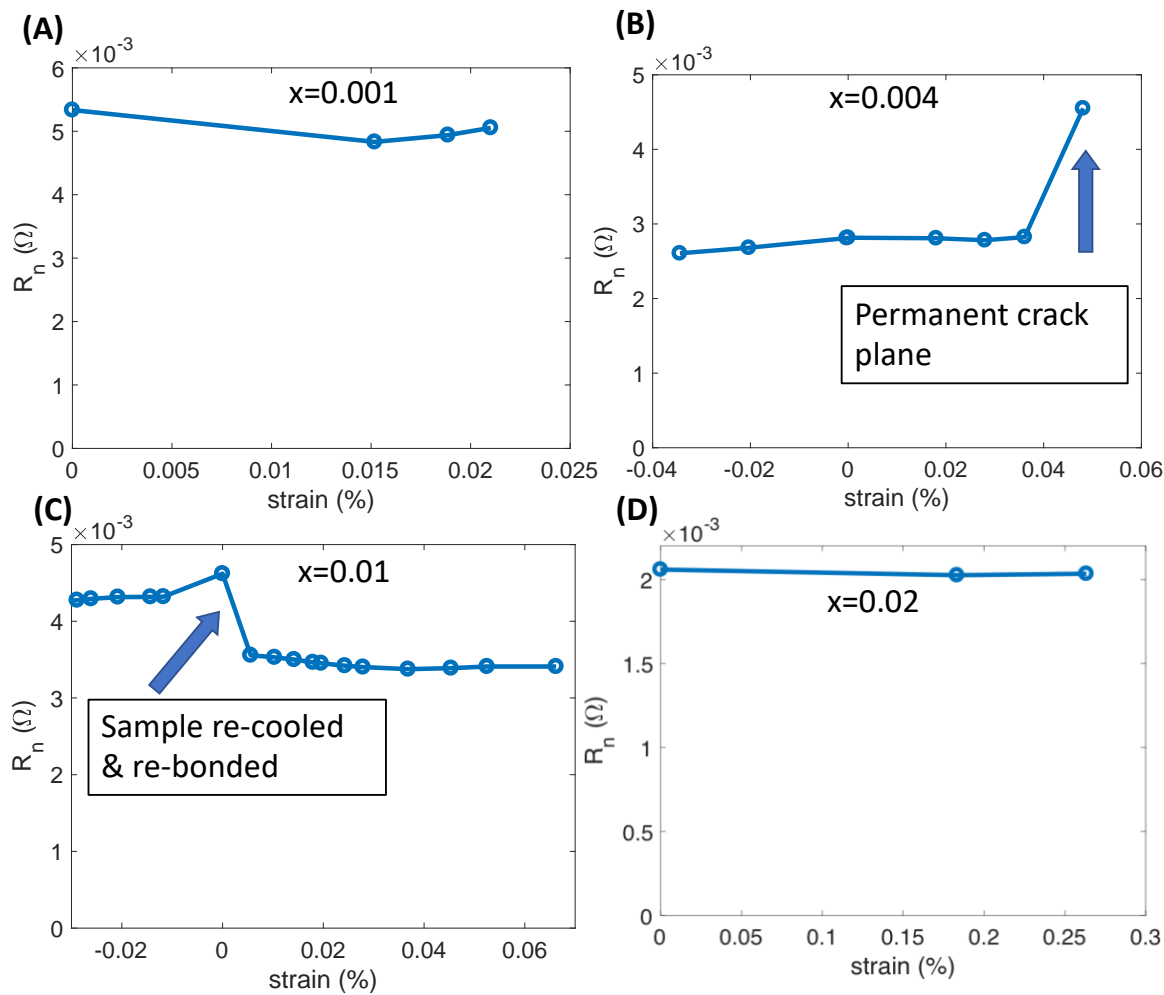

*Figure S 6. Normal state resistance for samples in Figure 4 (main text) shows no substantial changes under strain. (A) Sample resistance did not change substantially until the formation of a disconnected region in the sample (crack) that was observed optically. We still observed $T_c$ in the remaining connected part of the sample. (B) In this case, we repaired the broken sample using Stycast epoxy and observed no major change in the sample properties, besides those coming from repositioning the leads. Repairs required warming up to room temperature. Typical breaking tensile strains for x=0.01 and x=0.004 were slightly higher than 0.05%. For lower doping (x=0.001) the typical breaking tensile strain was about half the size. (C) Besides some small non-monotonous changes, no substantial change in the resistance in this sample was observed. (D) Insignificant change in the normal resistance of the overdoped sample. Overall, there is no substantial contribution from defects to the critical temperature changes (samples (A)-(D) in this figure are consistent with (A)-(D) in **Figure 4**).*

**Supplementary Materials References**